\documentclass[12pt]{article}
\usepackage{cite}
\usepackage{amssymb}
\usepackage{amsmath}
\usepackage[dvips]{graphicx,psfrag}

\renewcommand{\thefootnote}{\#\arabic{footnote}}
\newcommand{\slashp}{\not{\hbox{\kern-3pt $P$}}}
\newcommand{\slashs}{\not{\hbox{\kern-3pt $S$}}}

\begin{document}

\renewcommand{\thefootnote}{\fnsymbol{footnote}}
\setcounter{footnote}{0}
\begin{titlepage}

\def\thefootnote{\fnsymbol{footnote}}

\begin{center}

\hfill TU-666\\
\hfill CALT-68-2399\\
\hfill hep-ph/0208149\\
\hfill August, 2002\\

\vskip .5in

{\Large \bf
Top-Squark Study at a Future $e^+e^-$ Linear Collider
}

\vskip .45in

{\large
Ryuichiro Kitano$^{(a)}$, Takeo Moroi$^{(a)}$ and Shufang Su$^{(b)}$
}

\vskip .45in

{\em
$^{(a)}$Department of Physics, Tohoku University,
Sendai 980-8578, Japan
}
\vskip .2in

{\em
$^{(b)}$California Institute of Technology, 
Pasadena, CA 91125, USA
}

\end{center}

\vskip .4in

\begin{abstract}

    We discuss a potential of studying the production and the decay of
    the lightest top squark ($\tilde{t}_1$) in the framework of the
    supersymmetric standard model at a future $e^+e^-$ collider.  In
    particular, we consider the process $\tilde{t}_1\rightarrow
    t\chi^0_1$ (with $\chi^0_1$ being the lightest neutralino)
    followed by $t\rightarrow bW$.  It is shown that, by the study of
    the angular distribution of the bottom quark (as well as the
    production cross section of the top squark), properties of
    $\chi_1^0$ can be extracted.  We also discuss that, if $\chi^0_1$
    is gaugino-like, the neutralino mixing parameters (i.e.,\ the
    so-called $\mu$-parameter and $\tan \beta$) may be constrained.

\end{abstract}

\end{titlepage}

\renewcommand{\thepage}{\arabic{page}}
\setcounter{page}{1}
\renewcommand{\thefootnote}{\#\arabic{footnote}}
\setcounter{footnote}{0}

\section{Introduction}
\label{sec:intro} After the precise confirmation of the standard model
of the elementary particles by the LEP, SLD, and Tevatron, main purpose
of the high energy colliders are now shifting to discoveries and studies
of new physics beyond the standard model.  In particular, the Tevatron
Run II and LHC experiments are expected to give us strong hints for
studying the physics behind the electroweak symmetry breaking 
\cite{CDF,ATLAS}.  
As well as these hadron colliders, however, $e^+e^-$ linear
colliders are an alternative possibility to study the physics at the
electroweak scale.  Indeed, $e^+e^-$ linear colliders are considered as
serious candidates of the new program of experiments which pursue the
energy frontier of the high energy physics 
\cite{JLC1,NLC,JLC, snowmass,TESLA}.

Compared to the hadron colliders, $e^+e^-$ linear colliders may provide
cleaner environments to study productions and decays of various
particles with less background, which enables us to perform precision
measurements of the particle properties.  In addition, polarized $e^-$
beam will be available at $e^+e^-$ linear colliders which will be useful
to study the properties of the new particles.  Thus, $e^+e^-$ linear
collider will play a complementary role in studying the physics beyond
the standard model even if some signals of the new physics are
discovered at the Tevatron Run II and/or the LHC before the start of
$e^+e^-$ linear collider experiments.  Therefore, it is important to
understand what kind of information will be available from $e^+e^-$
linear colliders.

Among various possibilities, supersymmetry (SUSY) is one of the most
attractive candidates of the new physics and hence it is significant to
understand what can be studied at $e^+e^-$ linear colliders about SUSY
models.  The study is, however, not straightforward since there are many
parameters in the model and the collider phenomenology crucially depends
on these parameters.  In the simplest set up, capabilities of $e^+e^-$
linear colliders have been well studied \cite{Tsukamoto:1993gt}.  Most
of the studies have, however, assumed some simple mass spectra of the
superparticles.  For a more complete understanding of the potential of
$e^+e^-$ linear colliders, it is important to study various cases.

One of the possibilities is the case with a light stop (top squark).
For the stops, due to the large top Yukawa coupling constant, effects
of the left-right mixing in the mass matrix are expected to be large
and one of the stops can become light.  Thus, a light stop is easily
realized in wide parameter space and hence it is important to consider
what can be done with future high energy colliders.  Since the stops
affect the potential of the Higgs bosons through radiative corrections
\cite{Higgs1Loop}, stop study will provide useful information to
understand the physics of the electroweak symmetry breaking.  There
has been several works on the stop productions and decays at future
colliders.  At $e^+e^-$ linear colliders, the determination of the
mass and mixing of the stop from the stop-production cross section has
been discussed in detail in Ref.~\cite{Bartl:1997yi} for a 180 GeV
light stop by using the $\tilde{t}_1 \to \chi_1^0 c$ and $\tilde{t}_1
\to \chi^+ b$ modes. For the case of the LHC, it has been discussed
that the end-point analysis of gluino decay processes is useful to
understand properties of the stop \cite{ATLAS, Hisano:2002xq}.

In this paper, we consider the case where the lightest stop
$\tilde{t}_1$ is kinematically accessible at $e^+e^-$ linear colliders,
while it is still heavy enough to be able to decay into top quark and
the lightest neutralino $\chi^0_1$.  In particular, we consider the pair
production of the lightest stops and the decay of $\tilde{t}_1$ via
$\tilde{t}_1\rightarrow t\chi^0_1$ followed by $t\rightarrow bW$, and
discuss what can be studied.  It will be shown that the stop mass and
left-right mixing can be measured from the study of the production cross
section with polarized beam.  Also, the neutralino properties can be
extracted by the analysis of the stop decay.  The mass of the lightest
SUSY particle (LSP) is determined from the end-point of the decay
product of $\tilde{t}_1$.  In addition, information on the neutralino
mixing is obtained from the angular distribution of the $b$-jet.
Similar analysis using the measurements of the polarization of $\tau$
from stau decays is proposed in Ref.~\cite{Nojiri:1994it}.  The angular
distribution of the $b$-jet carries information on the top-quark
helicity. This fact enables us to extract neutralino mixing since the
helicity of the top quark from the stop decay is related to the
left-right stop mixing and the neutralino mixing effects, i.e., the
gauginos (superpartner of gauge boson) couple left- and right-handed
stops to left- and right-handed top quark, respectively, while the stop
and top quark have opposite helicities at the Higgsino (superpartner of
Higgs boson) vertices.  It is also discussed that, for gaugino-like LSP,
because of the large top Yukawa coupling constant, the angular
distribution of the $b$-jet may depend on so-called $\mu$-parameter
and hence our analysis provides some knowledge on $\mu$.

This paper is organized as follows.  In Sec.~\ref{sec:mass}, we
briefly introduce the stop sector and the neutralino sector, presenting
the mass and the mixing matrices.  In Sec.~\ref{sec:cross}, we
illustrate how to determine the lightest-stop mass and mixing angle
through the stop pair production with polarized electron beam.  In
Sec.~\ref{sec:decay}, we discuss in detail the decay mode
$\tilde{t}_1\rightarrow t \tilde\chi_1^0$, followed by $t \rightarrow
b W$.  The LSP mass can be obtained via the end-point of the energy
distribution of the top quark.  Sec.~\ref{sec:angular} is focusing on
the angular distribution of the $b$-jet from the top decay, based on
which we could obtain information on the neutralino mixing.  We study
three different neutralino LSP cases in Sec.~\ref{sec:mu}, and show
whether constrains on $\mu$ and $\tan\beta$ can be extracted.
Sec.~\ref{sec:conclusions} is devoted to conclusions.

\section{Top squark and neutralino sectors}
\label{sec:mass}

In the minimal supersymmetric standard model (MSSM), there are two
stops: $\tilde{t}_L$ and $\tilde{t}_R$, which are the superpartners of
the left- and right-handed top quarks, respectively. After the
electroweak symmetry breaking, these stops mix and become mass
eigenstates.  Mass matrix of the stops in the basis of $(\tilde{t}_L,
\tilde{t}_R)$ is given by
\begin{eqnarray}
    {\cal M}^2_{\tilde{t}} = 
    \left( \begin{array}{cc}
            m_{\tilde{t}L}^2 + m_t^2 + D_L
            & - y_t \mu \langle H_1\rangle 
            - A_t \langle H_2\rangle
            \\
            - y_t \mu \langle H_1\rangle 
            - A_t \langle H_2\rangle
            & m_{\tilde{t}R}^2 + m_t^2 + D_R
        \end{array} \right),
    \label{m2_st}
\end{eqnarray}
where $m_{\tilde{t}L}^2$ and $m_{\tilde{t}R}^2$ are the soft SUSY
breaking mass parameters for the left- and right-handed stops,
respectively, $y_t$ is the top Yukawa coupling constant, $m_t$ the top
quark mass, $\mu$ the SUSY invariant Higgs mass, and $A_t$ the
tri-linear coupling constant of stop.  For simplicity, we assume that
all the SUSY parameters are real in our analysis.  In addition, $H_1$
and $H_2$ are down- and up-type Higgs bosons, respectively, and
$\langle\cdots\rangle$ is the vacuum expectation value.  The $D$-term
contributions $D_L$ and $D_R$ are given by
\begin{eqnarray}
D_L = 
\left( \frac{g_2^2}{4} - \frac{g_1^2}{12} \right)
\left(\langle H_1 \rangle^2 - \langle H_2 \rangle^2 \right),~~~
D_R = \frac{g_1^2}{3}
\left(\langle H_1 \rangle^2 - \langle H_2 \rangle^2 \right),
\end{eqnarray}
where $g_1$ and $g_2$ are the gauge coupling constants of the U(1)$_Y$
and SU(2)$_L$ gauge interactions, respectively.

The stop mass matrix ${\cal M}^2_{\tilde{t}}$ can be diagonalized by
the unitary matrix $U_{\tilde{t}}$ to give mass eigenstates
$\tilde{t}_1$ and $\tilde{t}_2$:
\begin{eqnarray}
    \left( \begin{array}{c} \tilde{t}_1 \\
            \tilde{t}_2 \end{array} \right)
    =U_{\tilde{t}}\left( \begin{array}{c} 
            \tilde{t}_L \\ \tilde{t}_R \end{array} \right)
    =\left( \begin{array}{cc} 
            \cos\theta_{\tilde{t}} & \sin\theta_{\tilde{t}} \\
            - \sin\theta_{\tilde{t}} & \cos\theta_{\tilde{t}}
        \end{array} \right)
    \left( \begin{array}{c} 
            \tilde{t}_L \\ \tilde{t}_R \end{array} \right),
\end{eqnarray}
with mass eigenvalues $m_{\tilde{t}_1}^2$ and $m_{\tilde{t}_2}^2$.
$\theta_{\tilde{t}}$ is the mixing angle that parametrizes the
left-right stop mixing. (We define $m_{\tilde{t}_1}<m_{\tilde{t}_2}$ and
$-\frac{\pi}{2} \leq \theta_{\tilde{t}} \leq \frac{\pi}{2}$.)  With this
definition, the lightest stop becomes purely left-handed if
$\theta_{\tilde{t}}=0$ and right-handed if
$\theta_{\tilde{t}}=\frac{\pi}{2}$ and $-\frac{\pi}{2}$.  The left-right
mixing in the stop sector could be sizable due to the large top Yukawa
coupling constant.  Consequently, the mass of the lightest stop
$\tilde{t}_1$ could be so small that pair production of
$\tilde{t}_1\tilde{t}_1^*$ is kinematically allowed at $e^+e^-$ linear
colliders.

For the neutralino sector, Bino $\tilde{B}$, neutral Wino $\tilde{W}$,
neutral down-type Higgsino $\tilde{H}_1$, and neutral up-type Higgsino
$\tilde{H}_2$ are mixed together to form mass eigenstates. Mass matrix
of the neutralinos in the basis of $(\tilde{B}, \tilde{W},
\tilde{H}_1, \tilde{H}_2)$ is given by
\begin{eqnarray}
    {\cal M}_{\chi^0} = 
    \left( \begin{array}{cccc}
            M_1 & 0 & 
            \frac{1}{\sqrt{2}}g_1\langle H_1\rangle &
            - \frac{1}{\sqrt{2}}g_1\langle H_2\rangle \\
            0 & M_2 &
            - \frac{1}{\sqrt{2}}g_2\langle H_1\rangle &
            \frac{1}{\sqrt{2}}g_2\langle H_2\rangle \\
            \frac{1}{\sqrt{2}}g_1\langle H_1\rangle &
            - \frac{1}{\sqrt{2}}g_2\langle H_1\rangle &
            0 & - \mu \\
            - \frac{1}{\sqrt{2}}g_1\langle H_2\rangle &
            \frac{1}{\sqrt{2}}g_2\langle H_2\rangle &
            - \mu & 0
\end{array}\right),
\label{neutralino_mass_matrix}
\end{eqnarray}
where $M_1$ and $M_2$ are the U(1)$_Y$ and SU(2)$_L$ gaugino masses,
respectively.  The mass matrix ${\cal M}_{\chi^0}$ can be diagonalized
by a unitary matrix $U_{\chi^0}$, such that
\begin{equation}
{\cal M}_{\chi^0}^{\rm diag}=U_{\chi^0} {\cal M}_{\chi^0} U_{\chi^0}^T.
\end{equation}
The mass eigenstate of neutralino $\chi^0_i$ is thus given by 
\begin{eqnarray}
    \chi^0_i = 
    [U_{\chi^0}]^*_{i\tilde{B}} \tilde{B}
    + [U_{\chi^0}]^*_{i\tilde{W}} \tilde{W}
    + [U_{\chi^0}]^*_{i\tilde{H}_1} \tilde{H}_1
    + [U_{\chi^0}]^*_{i\tilde{H}_2} \tilde{H}_2,
\end{eqnarray}
with $\chi_1^0$ being the lightest neutralino (i.e., the LSP in our
analysis).  The neutralino mass eigenvalues $m_{\chi_i^0}$ and mixing
matrix $U_{\chi^0}$ are determined by four parameters $M_1$, $M_2$,
$\mu$ and $\tan\beta$.  Considering the case where $\langle H_1
\rangle, \langle H_1 \rangle \ll |M_1|, |M_2|, |\mu|$, the neutralino
mass eigenstates are nearly $\tilde{B}$, $\tilde{W}$, $(\tilde{H}_1
\pm \tilde{H}_2)/\sqrt{2}$, with mass eigenvalues $\sim$ $|M_1|$,
$|M_2|$ and $|\mu|$.  Depending on the relative magnitude of $|M_1|$,
$|M_2|$ and $|\mu|$, we would have Bino-like, Wino-like or
Higgsino-like lightest neutralino as the LSP.

As one can see, the mass matrices of the stops and the neutralinos
depend on various SUSY parameters.  Thus the mass spectrum of the
superparticles is really model dependent.  In this paper, to make our
points clear, we assume that the lightest stop $\tilde{t}_1$ is
accessible at $e^+e^-$ linear colliders but the heavier one
$\tilde{t}_2$ is not.\footnote
{When $m_{\tilde{t}_1}+m_{\tilde{t}_2}< E_{\rm cm}$, the associate
production $e^+e^-\rightarrow\tilde{t}_1\tilde{t}_2^* $ would give
us information on the heavy stop.  However, we will not discuss such
case in this paper.}

\section{Stop pair production}
\label{sec:cross}

Let us first consider the pair production process of the lightest
stop.  Importantly, it is expected that the electron beam can be
polarized at future $e^+e^-$ linear colliders, which is extremely
useful for the study of the stops \cite{Bartl:1997yi}. The production
cross section with the left-polarized electron beam is given by
\begin{eqnarray}
\sigma (e^+e_L^-\rightarrow\tilde{t}_1\tilde{t}^*_1) =
\frac{1}{16\pi s} 
\left( Q_eQ_t + g^{e_L}_Zg^{\tilde{t}_1}_Z
\frac{s}{s-m_Z^2} \right)^2
\left( 1 - \frac{4m_{\tilde{t}_1}^2}{s} \right)^{3/2},
\label{eq:cross}
\end{eqnarray}
where $\sqrt{s}$ is the center of mass energy of the linear collider.
Similar formula for $\sigma
(e^+e_R^-\rightarrow\tilde{t}_1\tilde{t}^*_1)$ is obtained by replacing
$L\rightarrow R$. The charges and couplings in Eq.~(\ref{eq:cross}) are
given by
\begin{eqnarray}
Q_e = -e,~~~
g^{e_L}_Z = \frac{- g_2^2 +  g_1^2}{2g_Z},~~~
g^{e_R}_Z = \frac{g_1^2}{g_Z},
\end{eqnarray}
and
\begin{eqnarray}
Q_t = \frac{2}{3} e,~~~
g^{\tilde{t}_1}_Z =
\frac{1}{g_Z} \left( \frac{1}{2} g_2^2 - \frac{1}{6} g_1^2 \right)
\cos^2 \theta_{\tilde{t}}
- \frac{2g_1^2}{3g_Z} \sin^2 \theta_{\tilde{t}},
\end{eqnarray}
where $g_Z=\sqrt{g_2^2+g_1^2}$, $e=g_1g_2/g_Z$, and $m_Z$ is the
$Z$-boson mass.

Diagrams with photon-exchange and $Z$-exchange both contribute to the
production cross section, which interfere either constructively or
destructively depending on the polarization of the electron beam and
the stop mixing angle.
\begin{figure}[tp]
\hspace*{0.5cm}
\includegraphics[height=7cm]{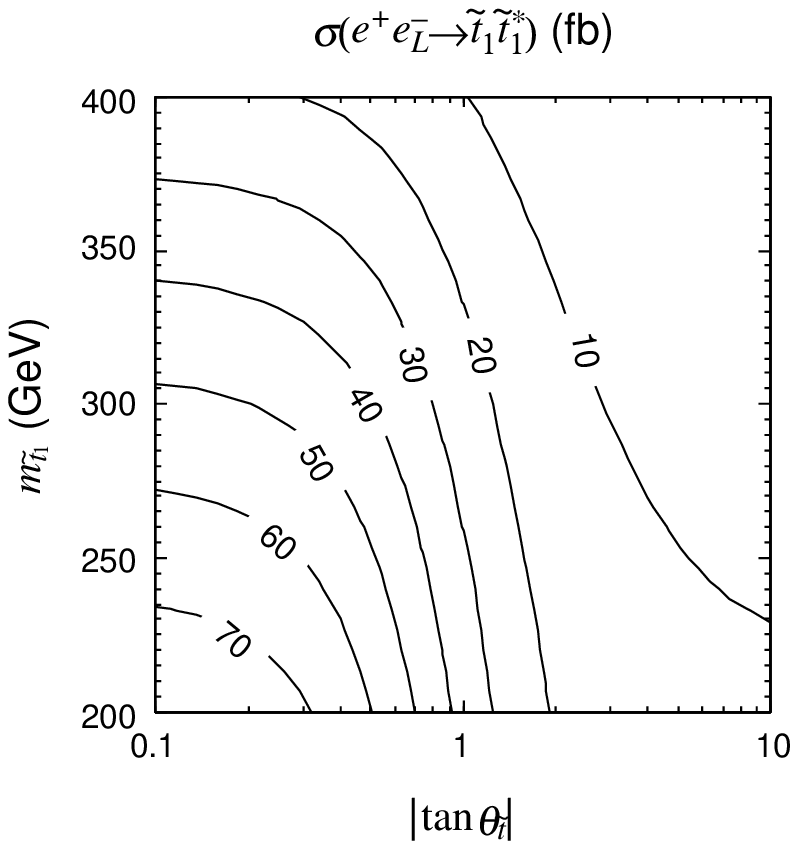}
\hspace*{5mm}
\includegraphics[height=7cm]{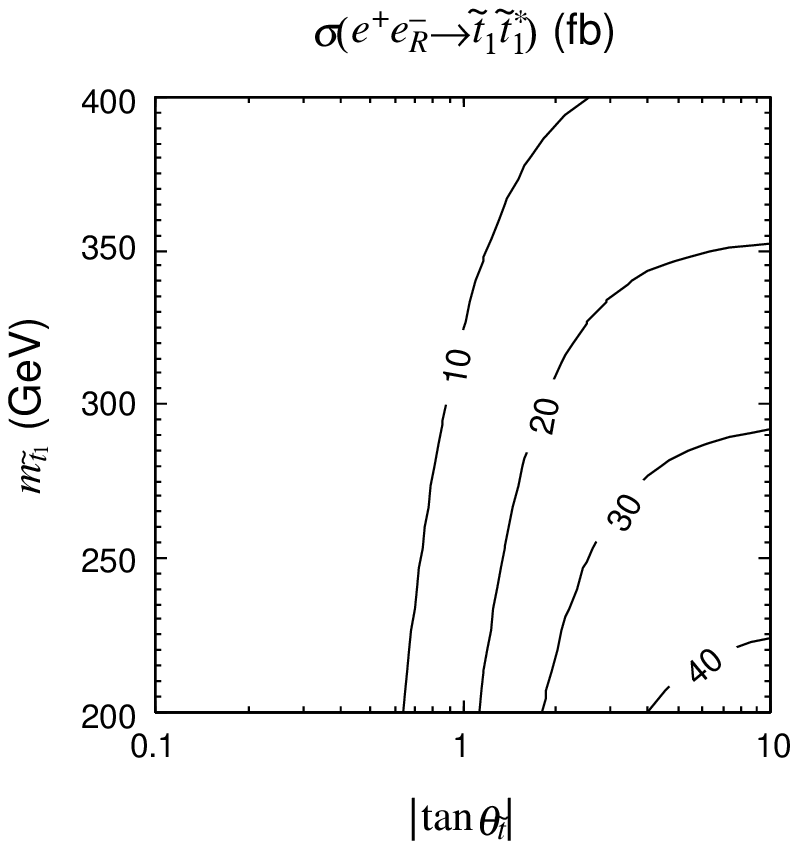}  
\caption{Contours of constant $\sigma
(e^+e^-_L\rightarrow\tilde{t}_1\tilde{t}^*_1)$ (left plot) and $\sigma
(e^+e^-_R\rightarrow\tilde{t}_1\tilde{t}^*_1)$ (right plot) on the
$m_{\tilde{t}_1}$ vs.\ $|\tan \theta_{\tilde{t}}|$ plane at a linear 
collider with center-of-mass energy $\sqrt{s}=1\ {\rm TeV}$.}
\label{fig:sigma}
\end{figure}

In Fig.\ \ref{fig:sigma}, we plot the contours of constant $\sigma
(e^+e^-_L\rightarrow\tilde{t}_1\tilde{t}^*_1)$ and $\sigma
(e^+e^-_R\rightarrow\tilde{t}_1\tilde{t}^*_1)$ on the
$m_{\tilde{t}_1}$ vs.\ $|\tan \theta_{\tilde{t}}|$ plane.  (Here and
hereafter, we take $\sqrt{s}=1\ {\rm TeV}$.)  The cross sections are
sensitive to both the mass and mixing of the stop, as expected.  
The behaviors of the cross sections can be understood in the following
way.  With left-polarized electron beam, photon and $Z$-exchange
diagrams interfere constructively when the lightest stop is almost
left-handed (i.e., when $|\tan\theta_{\tilde{t}}|$ is small), and
destructively when the lightest stop is almost right-handed.  Such
effect is shown in the left plot of Fig.\ \ref{fig:sigma}; the cross
section decreases as $|\tan\theta_{\tilde{t}}|$ increases.  With
right-polarized beam, the interference term flips sign.  Therefore,
the cross section increases as $|\tan \theta_{\tilde{t}}|$ increases,
which is shown in the right plot of Fig.\ \ref{fig:sigma}.

Most importantly, both $m_{\tilde{t}_1}$ and $\theta_{\tilde{t}}$ can
be determined simultaneously using the cross-section measurement with
polarized beam \cite{Bartl:1997yi,PRD49-2369,LCstopstudy} since
dependences of $\sigma (e^+e^-_L\rightarrow\tilde{t}_1\tilde{t}^*_1)$
and $\sigma (e^+e^-_R\rightarrow\tilde{t}_1\tilde{t}^*_1)$ on
$m_{\tilde{t}_1}$ and $\theta_{\tilde{t}}$ are significantly
different. (See Fig.\ \ref{fig:sigma}.)

In the following analysis, we use the averaged cross section in
estimating the number of events:
\begin{eqnarray}
\sigma = \frac{1}{2} \left[ 
\sigma(e^+e^-_L\rightarrow\tilde{t}_1\tilde{t}^*_1) +
\sigma(e^+e^-_R\rightarrow\tilde{t}_1\tilde{t}^*_1)
\right].
\end{eqnarray}

\section{Stop decay}
\label{sec:decay}

Given that we have known the value of $m_{\tilde{t}_1}$ and
$\theta_{\tilde{t}}$ from the measurements of the stop production
cross sections, we now study whether the stop decay renders additional
information on the stop and neutralino sectors.  In our analysis, we
use the decay mode $\tilde{t}_1\rightarrow t\chi^0_1$ followed by
$t\rightarrow bW$.  Reconstruction of the top quark events provides us
energy distribution of the top quark and angular distribution of the
$b$-jet.  The former may be used to obtain the mass of the LSP.  In
addition, the latter is sensitive to helicity of the top quark which
depends on the LSP properties.  Therefore, the angular distribution of
the $b$-jet can be used to extract informations on the LSP.

Classifying the decay modes of the $W$-boson, the possible final
states of the $\tilde{t}_1\tilde{t}^*_1$ production process are:
\begin{itemize}
\item[(a)] $2b$ + $4q$ + missing momentum (48\ \%),
\item[(b)] $2b$ + $2q$ + $1l$ + missing momentum (43\ \%),
\item[(c)] $2b$ + $2l$ + missing momentum (9\ \%),
\end{itemize}
where $b$ and $q$ are hadronic jets with and without a $b$-tag,
respectively, and $l$ is the charged lepton.\footnote
{We assume that we can identify the $\tau$ event even if $\tau$ decays
into hadrons since the multiplicity of the hadronic decay product from
$\tau$ is very small.  Thus, we treat the $\tau$ lepton as other
charged leptons.}
With the vertex detector, we expect that the $b$-jet will be
distinguished from other hadronic activities with high efficiency.  In
order to carry out the analysis of the stop decay, at least one of the
$W$-boson should decay hadronically so that the top quark can be
reconstructed.  Thus, the events with final states (a) and (b) are
used for the following analysis.

We assume that there is no SUSY background, and we expect the
standard-model backgrounds from the following processes
\cite{ut580}:\footnote
{The process $e^+e^-\rightarrow t\bar{t}$, followed by the decays $t
\rightarrow b W$, $W^\pm\rightarrow l\bar{\nu}$, and $W^\mp\rightarrow
{\rm hadrons}$, has the same final state as the signal event.  In this
process, however, the energy of the top quark is $\sqrt{s}/2$.  In addition,
the invariant mass constructed from the missing energy and momentum
vanishes.  Thus, we assume that this background can be eliminated by
imposing cuts on these variables.}
\begin{itemize}
\item[(i)] $t\bar{t}Z$ followed by $Z\rightarrow\nu\bar{\nu}$
($\sim 1\ {\rm fb}$)
\item[(ii)] $t\bar{t}\nu_e\bar{\nu}_e$ ($\sim 0.5\ {\rm fb}$)
\item[(iii)] $ZW^\pm W^\mp$ followed by 
$Z\rightarrow b\bar{b}$, $W^\pm\rightarrow l\bar{\nu}$, and
$W^\mp\rightarrow {\rm hadrons}$
($\sim 3\ {\rm fb}$) 
\item[(iv)] $e^{\pm}\nu_e(\bar{\nu}_e)W^\mp Z$ 
followed by $Z\rightarrow b\bar{b}$, and
$W\rightarrow {\rm hadrons}$
($\sim 11\ {\rm fb}$)
\end{itemize}
We assume that the backgrounds (iii) and (iv) can be eliminated by
imposing relevant cut since the invariant mass of two $b$-jets is
equal to $m_Z$.  On the contrary, the processes (i) and (ii) provide
irreducible backgrounds.  These are, however, standard-model processes
and hence their cross sections are calculable.  Therefore, we can
subtract the number of backgrounds from the number of signal
candidates in the measurements of the cross sections.  In addition,
the cross sections for these processes are order-of-magnitude smaller
than the signal cross section.  Thus, as we will see later, effects of
these backgrounds are insignificant.

In a stop pair production event, if all the lightest stop decays into
$t$ and $\chi_1^0$, the averaged number of stops available for the
analysis (process (a) and (b)) is
\begin{eqnarray}
\epsilon = 2 \epsilon_b^2 Br(W\rightarrow {\rm had.}),
\label{eq:eff}
\end{eqnarray}
where $\epsilon_b$ is the efficiency of the $b$-tagging and
$Br(W\rightarrow {\rm had.})\simeq 0.69$ is the hadronic branching
ratio of the $W$-boson. With $\epsilon_b=60\ \%$ \cite{snowmass}, we
use $\epsilon\simeq 0.50$.  Notice that if additional stop decay modes
are available, for example, $\tilde{t}_1 \rightarrow b \chi_1^{\pm}$
in Wino-LSP and Higgsino-LSP case, $\epsilon$ should be multiplied by
an additional factor of $Br(\tilde{t}_1\rightarrow
t\chi_1^0)$.\footnote
{In certain cases, the additional decay modes of the lightest stop
result in the same final state as the signal and become SUSY
background.  Analysis in such a situation is beyond the scope of the
current study.}
In this case, measurement of the branching ratio of $\tilde{t}_1$
provides additional information on the neutralino/chargino sector.

Once the top quark from the decay of $\tilde{t}_1$ is reconstructed,
the mass of the LSP can be determined by measuring the energy
distribution of the top quark.  Indeed, the maximum and minimum energy
of the top quark in the lab frame, $E_t^{\rm (max)}$ and $E_t^{\rm
(min)}$, respectively, are given by
\begin{eqnarray}
    E_t^{\rm (max)} = 
    \frac{E_t^* + p_t^* \beta_{\tilde{t}_1}}
    {\sqrt{1-\beta_{\tilde{t}_1}^2}},~~~
    E_t^{\rm (min)} = 
    \frac{E_t^* - p_t^* \beta_{\tilde{t}_1}}
    {\sqrt{1-\beta_{\tilde{t}_1}^2}},
\end{eqnarray}
where
\begin{eqnarray}
    E_t^* = \frac{m_{\tilde{t}_1}^2 - m_{\chi^0_1}^2 + m_t^2}
    {2m_{\tilde{t}_1}},~~~
    p_t^* = \sqrt{(E_t^*)^2 - m_t^2},
\end{eqnarray}
and $\beta_{\tilde{t}_1}=(1-4m_{\tilde{t}_1}^2/s)^{1/2}$.  By
measuring these end-points, we obtain information about the mass of
the LSP (as well as that about the stop mass) \cite{PRD49-2369}.  At
this stage, the parameters $m_{\tilde{t}_1}$, $\theta_{\tilde{t}}$,
and $m_{\chi^0_1}$ are determined from the stop production cross
sections and stop decay kinematics.  In the following analysis, we
assume that these three parameters are well understood.

\section{Angular distribution of $b$-jet}
\label{sec:angular}

Stop couples to the gauginos and to up-type Higgsino through the gauge
and Yukawa interactions, respectively.  One striking difference
between the couplings to the gauginos and that to the Higgsino is that
the left-handed (right-handed) stop couples {\sl only} to left-handed
(right-handed) top quark via gaugino interactions, while the helicity
is flipped via the Higgsino interaction.  Therefore, information about
the stop and the LSP is imprinted in the helicity of the top quark and
hence the measurement of the helicity of the top quark provides some
knowledge of these particles.  This point becomes clearer by looking
at the top-stop-neutralino vertex.  The coupling of the lightest stop
$\tilde{t}_1$ with the top and the lightest neutralino is given by
\begin{eqnarray}
    {\cal L}_{\rm int} &=& \tilde{t}_1^* 
    \bar{\chi}^0_1 ( f_L P_L + f_R P_R ) t 
    + \tilde{t}_1
    \bar{t} ( f_L^* P_R + f_R^* P_L ) \chi^0_1,
    \label{vertex}
\end{eqnarray}
where $P_{L/R}=\frac{1}{2}(1\mp\gamma_5)$, respectively, and 
\begin{eqnarray}
    f_L &=& 
    \left[ \frac{1}{\sqrt{2}} g_2 [ U_{\chi^0} ]_{1 \tilde{W}}
        + \frac{1}{3\sqrt{2}} g_1 [ U_{\chi^0} ]_{1 \tilde{B}}
    \right] \cos\theta_{\tilde{t}}
    - y_t [ U_{\chi^0} ]_{1 \tilde{H}_2} \sin\theta_{\tilde{t}},
\label{f_L}
    \\
    f_R &=& - \frac{2\sqrt{2}}{3} g_1
    [ U_{\chi^0} ]^*_{1 \tilde{B}} \sin\theta_{\tilde{t}}
    - y_t [ U_{\chi^0} ]^*_{1 \tilde{H}_2} \cos\theta_{\tilde{t}}.
\label{f_R}
\end{eqnarray}
Thus, for the case where the lightest stop is $\tilde{t}_L$, for
example, the top quark becomes left-handed if $\chi^0_1$ is
gaugino-like while it becomes right-handed if $\chi^0_1$ is
Higgsino-like.

In order to measure the helicity of the top quark, it is efficient to
study the angular distribution of decay products from the top, like
the $b$-quark \cite{kane}. 
The angular distribution of $b$-jet from the decay of a top quark with 
certain helicity can be obtained via direct computation, or by using the 
spin vector method.  Here we use the latter.
With Eq.\ (\ref{vertex}), the spin vector $S^\mu$ for
the top quark in the decay process $\tilde{t}_1\rightarrow t\chi^0_1$
is given by
\begin{eqnarray}
    &&\frac{1}{2} N (\slashp_t + m_t) (1 + \slashs\gamma_5)
     \nonumber \\ 
    &&\equiv
    (\slashp_t + m_t) ( f_L^* P_R + f_R^* P_L ) 
    (\slashp_\chi - m_{\chi^0_1}) 
    ( f_L P_L + f_R P_R ) (\slashp_t + m_t),
\end{eqnarray}
with $P_t$ and $P_\chi$ being four-momenta of the top quark and the LSP,
respectively.  Then,
\begin{eqnarray}
    N &=& (m_{\tilde{t}_1}^2 - m_t^2 - m_{\chi^0_1}^2) (|f_L|^2 + |f_R|^2)
    - 4 m_t m_{\chi^0_1} {\rm Re} ( f_L^* f_R ),
    \\
    N S^\mu &=& - \frac{1}{m_t} \left[ 2 m_t^2 P_\chi^\mu 
        - (m_{\tilde{t}_1}^2 - m_t^2 - m_{\chi^0_1}^2) P_t^\mu \right]
    (|f_L|^2 - |f_R|^2).
\end{eqnarray}

For the process $\tilde{t}_1\rightarrow t\chi^0_1$ followed by
$t\rightarrow bW$, the angular distribution of the $b$-jet {\sl in the
rest frame of the top quark} is given by
\begin{eqnarray}
    \frac{1}{\Gamma_{\tilde{t}}}
    \frac{d\Gamma_{\tilde{t}}}{d\cos\theta_{tb}}
    \equiv
    \frac{1}{2} ( 1 + A_b \cos\theta_{tb} ),
\end{eqnarray} 
where $\theta_{tb}$ is the angle between top and bottom quark jets in
the rest frame of the top quark,\footnote{
The opening angle $\theta_{tb}$ in the rest frame of the top quark is
given by the relation $\cos\theta_{tb}=(\cos\theta_{tb}^\prime-\beta)/
(1-\beta\cos\theta_{tb}^\prime)$, where $\theta_{tb}^\prime$ and $\beta$
are the opening angle and the velocity of the top quark in the
lab-frame, respectively. Alternatively, we can obtain the angle by the
energies of the top and the bottom quarks in the lab-frame by
$\cos\theta_{tb}=(E_b/E_b^{\rm rest} - \gamma)/ \gamma\beta$, where
$E_b$ and $E_b^{\rm rest}$ are the bottom energy in the lab-frame and in
the rest frame of the top quark, respectively, and $\gamma =
1/\sqrt{1-\beta^2}$. }
and the coefficient $A_b$ is related to the spin vector $S^\mu$ via 
\begin{eqnarray}
    A_b = - \frac{m_t^2 - 2m_W^2}{m_t^2 + 2m_W^2} \hat{S}_t,  
    \label{A_b}
\end{eqnarray} 
with
\begin{eqnarray}
    \hat{S}_t &=& - \frac{2m_t|{\bf P_\chi}|}{N}(|f_L|^2 - |f_R|^2)
    \nonumber \\ &=&
    - \frac{[m_{\tilde{t}_1}^2-(m_t-m_{\chi^0_1})^2]^{1/2}
    [m_{\tilde{t}_1}^2-(m_t+m_{\chi^0_1})^2]^{1/2}}
    {(m_{\tilde{t}_1}^2 - m_t^2 - m_{\chi^0_1}^2) (|f_L|^2 + |f_R|^2)
    - 4 m_t m_{\chi^0_1} {\rm Re} ( f_L^* f_R )} (|f_L|^2 - |f_R|^2).
    \label{S-hat}
\end{eqnarray}
Here ${\bf P}_\chi$ is the three-momentum of the LSP in the rest
frame of the top quark and $\hat{S}_t^2=-S^\mu S^\mu$.  Notice that
$A_b$ depends only on $f_L/f_R$ and there is a two-fold degeneracy.

We can also obtain the angular distribution of the $s$-jets assuming
that the $W$-boson decays into $c$ and $s$:
\begin{eqnarray}
    \frac{1}{\Gamma_{\tilde{t}}}
    \frac{d\Gamma_{\tilde{t}}}{d\cos\theta_{ts}}
    \equiv
    \frac{1}{2} ( 1 + A_{s}\cos\theta_{ts} ),
\end{eqnarray} 
with 
\begin{eqnarray}
    A_{s} = \hat{S}_t.
\end{eqnarray}
In this case, there is no suppression factor which appears in
Eq.(\ref{A_b}).  However, it is difficult to directly identify the
$s$-jet, although one possibility may be to tag the $c$-jet as well as
$b$-jet.  Thus, we believe it is challenging to measure $A_s$ and, in
the following, we concentrate on the analysis of $A_b$.

\begin{figure}[tp]
\begin{center}
\includegraphics[width=8cm]{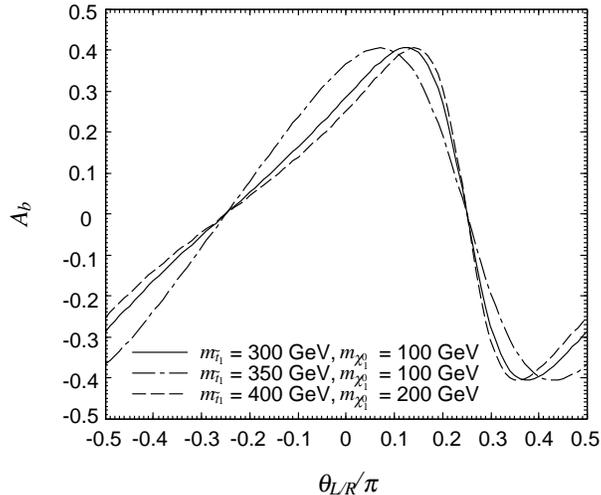}  
\caption{$A_b$ as a function of $\tan\theta_{L/R}=f_L/f_R$.  Here, we
take $\sqrt{s}=1\ {\rm TeV}$, and 
the stop and neutralino masses are taken to be
$ m_{\tilde{t}_1}= 300$ GeV and  
$m_{\chi^0_1} = 100$ GeV (solid line),
$m_{\tilde{t}_1}= 350$ GeV and  
$m_{\chi^0_1} = 100$ GeV (dash-dotted line), and
$m_{\tilde{t}_1}= 400$ GeV and  
$m_{\chi^0_1} = 200$ GeV (dashed line).
}
\label{fig:theta_vs_A}
\end{center}
\end{figure}

Given the fact that it is difficult to measure the total decay width
of $\tilde{t}_1$, the only information is given in the form of the
angular distribution of the bottom quark from the decay chain of
$\tilde{t}_1$.  Even so, the ratio $f_L/f_R$ can be, in principle,
determined and hence we obtain the information about the properties of
the stop and the LSP.  Indeed, the $A_b$-parameter has strong
dependence on the ratio $f_L/f_R$.  In Fig.\ \ref{fig:theta_vs_A}, we
plot $A_b$ as a function of $f_L/f_R$ for several values of
$m_{\tilde{t}_1}$ and $m_{\chi^0_1}$.  Here, we define the angle
$\theta_{L/R}$ as
\begin{eqnarray}
\tan\theta_{L/R} = f_L/f_R .
\end{eqnarray}
$A_b$ is asymmetric with respect to the sign of $\theta_{L/R}$ because
of the ${\rm Re}( f_L^* f_R )$ term in the denominator of Eq.\
(\ref{S-hat}).  As one can see, $A_b$ changes from about $-0.4$ to $0.4$
as $\theta_{L/R}$ varies; the steepest dependence on $\theta_{L/R}$
occurs around $\theta_{L/R}\sim \pi/4 $ (i.e., $f_L/f_R\sim 1$).  Thus,
measurement of the $A_b$-parameter provides constraint on the ratio
$f_L/f_R$ although, unfortunately, the two-fold ambiguity always
remains.

In order to discuss the statistical uncertainties in the measurement
of $A_b$, we express $A_b$ in terms of numbers of signal events
$N_{\rm S}^\pm$, where the superscripts ``$+$'' and ``$-$'' are for
the events with $\cos\theta_{tb}>0$ and $\cos\theta_{tb}<0$,
respectively: $N_{\rm S}^\pm \propto 1/2 \pm A_b/4$.  Similarly, we
define number of the background events as $N_{\rm B}^\pm$.  With these
variables, total number of events are given by $N_{\rm tot}^\pm=N_{\rm
S}^\pm +N_{\rm B}^\pm$, and the $A_b$-parameter is expressed as
\begin{eqnarray}
A_b = 
\frac{ 2 (N_{\rm S}^+ - N_{\rm S}^-) }
{N_{\rm S}^+ + N_{\rm S}^-}
=
\frac{2(N_{\rm tot}^+ - N_{\rm tot}^-) 
- 2(N_{\rm B}^+ - N_{\rm B}^-)}
{N_{\rm tot}-N_{\rm B}},
\label{Abformula}
\end{eqnarray}
where $N_{\rm tot}=N_{\rm tot}^++N_{\rm tot}^-$ and $N_{\rm B}=N_{\rm
B}^++N_{\rm B}^-$.  Notice that $N_{\rm B}^\pm$ are calculable quantities.
Then, the statistical uncertainty in the measurement of $A_b$ is
estimated to be
\begin{eqnarray}
(\delta A_b)^2 = \frac{16[N_{\rm tot}^+(N_{\rm tot}^--N_{\rm B}^-)^2+
N_{\rm tot}^-(N_{\rm tot}^+-N_{\rm B}^+)^2]}
{(N_{\rm tot}-N_{\rm B})^4}
\simeq 
\frac{4}{N_{\rm tot}},
\label{dAb}
\end{eqnarray}
where, in the second equality, we use the relations $N_{\rm B}\ll
N_{\rm tot}$ and $A_b^2\ll 1$.  Without these simplifications,
$(\delta A_b)^2$ increases about 10\ \%.\footnote{
The error from the uncertainty in $\cos\theta_{tb}$ can be safely
neglected.  To discuss this issue, it is convenient to use the fact that
$A_b$ is experimentally determined by using the following formula
instead of Eq.\ (\ref{Abformula}):
\begin{eqnarray*}
    A_b = \frac{3}{N_S} \sum_i \cos\theta_{tb,i},
\end{eqnarray*}
where the sum is over the signal events and $\theta_{tb,i}$ is
$\theta_{tb}$ for $i$-th event. $N_S$ is the numbers of the signal
events.  Then, the error in $A_b$ associated with the uncertainty in
$\cos\theta_{tb}$ is estimated as
\begin{eqnarray*}
    (\delta A_b)^2 = \frac{9}{N_S} (\delta\cos\theta_{tb})^2,
\end{eqnarray*}
where $\delta\cos\theta_{tb}$ is the error in $\cos\theta_{tb}$.
(Here, we neglect backgrounds to discuss effects of
$\delta\cos\theta_{tb}$.)  $\delta\cos\theta_{tb}$ is estimated using
the following formula:
\begin{eqnarray*}
    \cos\theta_{tb} = 
    \frac{ (E_b / E_b^{\rm rest}) - (E_t / m_t)}
    {\sqrt{(E_t/m_t)^2 - 1}},
\end{eqnarray*}
where $E_b$ and $E_t$ are energies of the bottom and top quarks in the
laboratory frame, respectively, and $E_b^{\rm rest}$ is the bottom-quark
energy in the rest frame of the top quark.  Imposing relevant
kinematical constraints, $E_b$ and $E_W\equiv E_t-E_b$ are determined at
the linear collider with uncertainties of about 5\ GeV and 3\ GeV,
respectively \cite{Ikematsu:2002zd}.  With such uncertainties,
$\delta\cos\theta_{tb}$ is estimated to be at most 0.15.  Thus, the
error in $A_b$ induced from $\delta\cos\theta_{tb}$ is much smaller than
the statistical error.
}

Constraints on the $A_b$-parameter can be translated to the bound on the
ratio $f_L/f_R$.  In order to estimate the expected bound on $f_L/f_R$
for a certain underlying value of $f_L/f_R$, we first fix the masses
($m_{\tilde{t}_1}$ and $m_{\chi^0_1}$) and the stop mixing angle
$\theta_{\tilde{t}}$.  Then, we determine $A_b$ with Eqs.\ (\ref{A_b})
and (\ref{S-hat}) by inputting the underlying value of $f_L/f_R$.
Simultaneously, with neglecting the backgrounds, we calculate the total
number of events $N_{\rm tot}=\epsilon L\sigma$ for a given integrated
luminosity $L$ and estimate the uncertainty of $A_b$ using Eq.\
(\ref{dAb}).  Then, to put the bound, we postulate the hypothetical
value of $f_L/f_R$ and calculate $A_b$, and determine if such a value is
within the experimental bound on $A_b$.

\begin{figure}[tp]
\hspace*{0.5cm}
\includegraphics[width=9.3cm]{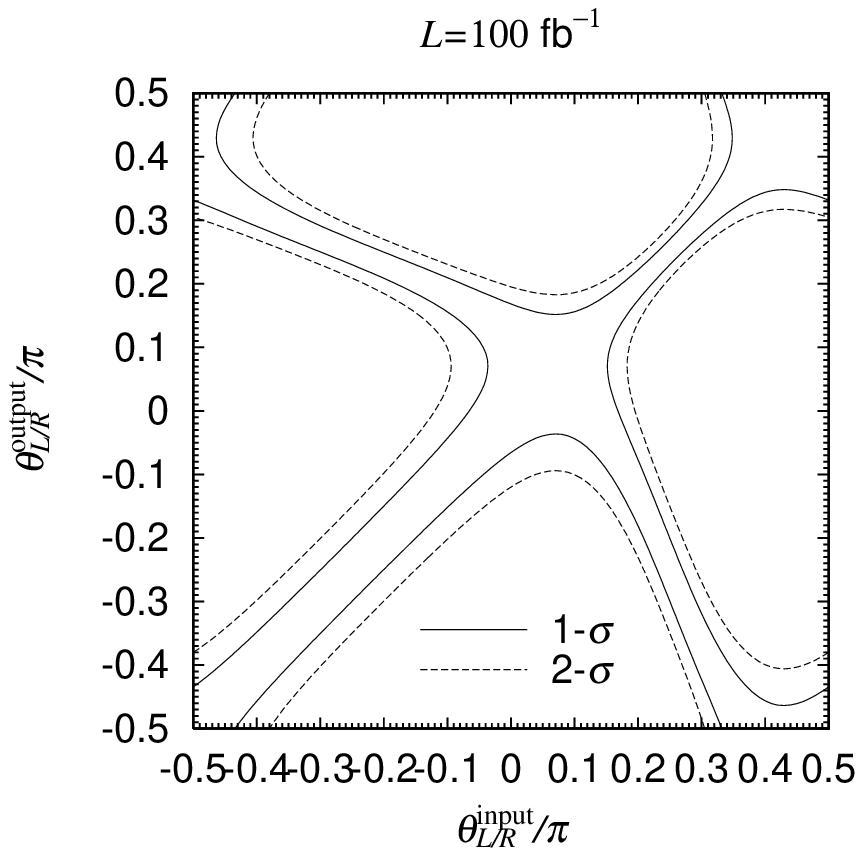}  
\hspace*{-2cm}
\includegraphics[width=9.3cm]{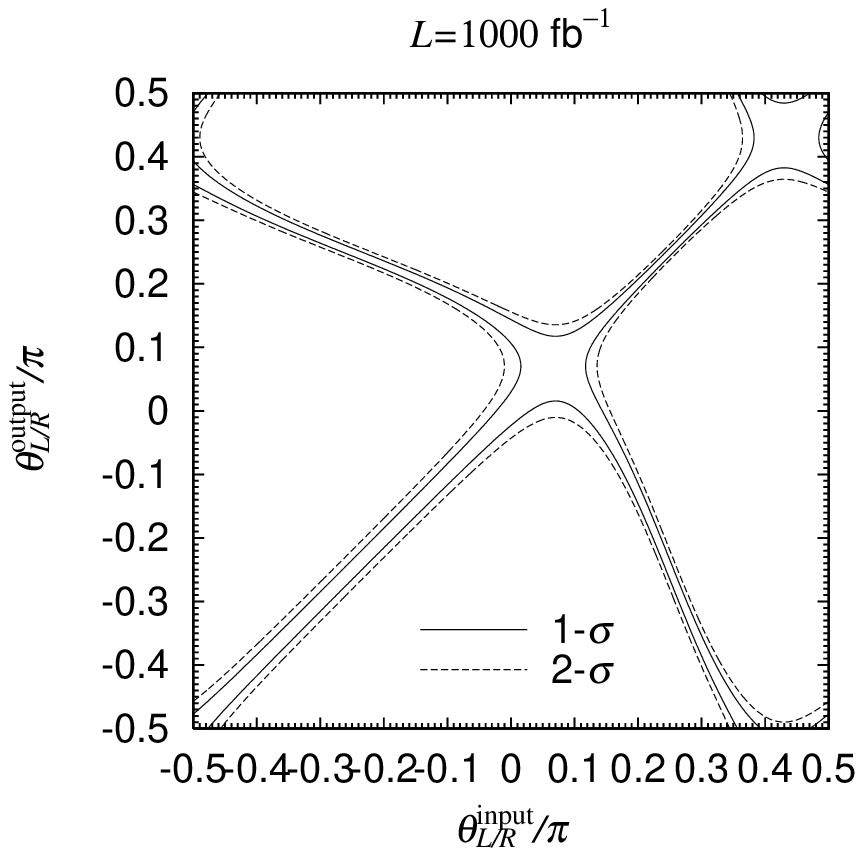}  
\caption{Expected upper and lower bounds at 1-$\sigma$ (solid line) 
and 2-$\sigma$ (dashed line) level on $\theta_{L/R}$ as a function of
the input-value of $\theta_{L/R}$.  We take $m_{\tilde{t}_1} = 350\ 
{\rm GeV}$, $\theta_{\tilde{t}}=\pi/4$, and $m_{\chi^0_1} = 100\ {\rm
GeV}$.  The total numbers of events $N_{\rm tot}$ are set to be $650$
and $6500$ corresponding to $L=100$ fb$^{-1}$ (left plot) and $1000$
fb$^{-1}$ (right plot), respectively.}
\label{fig:in_vs_out}
\end{figure}

In Fig.\ \ref{fig:in_vs_out}, we plot the expected upper and lower
bounds on $\theta_{L/R}$ as a function of the input value of
$\theta_{L/R}$.  Here, we take $m_{\tilde{t}_1} = 350\ {\rm GeV}$,
$\theta_{\tilde{t}}=\pi/4$, $m_{\chi^0_1} = 100\ {\rm GeV}$, and $L=100\
{\rm fb}^{-1}$ (left plot) and $L=1000\ {\rm fb}^{-1}$ (right plot).
With these parameters, the total number of events is given by $N_{\rm
tot} \simeq 650$ ($L=100\ {\rm fb}^{-1}$) and 6500 ($L=1000\ {\rm
fb}^{-1}$), which give $\delta A_b\simeq 0.08$ and 0.02, respectively.

For one $\theta_{L/R}^{\rm input}$, there are two possible regions of
$\theta_{L/R}^{\rm output}$, due to the two-fold ambiguities that we
mentioned earlier.  For $\theta_{L/R}^{\rm input}$ close to 0 and
$\pi/2$, which corresponds to the region where $A_b$ is at its minimum
or maximum, two regions merge.  In the case of $\theta_{L/R}>0$,
higher precision is obtained compared to the case of $\theta_{L/R}<0$
since the $A_b$ dependence on $\theta_{L/R}$ becomes steeper when
$\theta_{L/R}$ is positive.  With $L=100\ {\rm fb}^{-1}$, the
uncertainty of $\theta_{L/R}/\pi$ can be as small as 0.05 (0.1) at
1-$\sigma$ (2-$\sigma$) although the two-fold ambiguity exists.

\section{$\mu$-parameter}\label{sec:mu}

Since the parameters $m_{\tilde{t}_1}$, $\theta_{\tilde{t}}$, and
$m_{\chi^0_1}$ are determined without information about $A_b$, the
measurement of $\theta_{L/R}$, which depends on the neutralino mixing
matrix $U_{\chi^0}$, can be used to study the neutralino sector.
However, information we could obtain depends on what the lightest
neutralino is.  So, we consider the three typical cases: Bino-like
LSP, Wino-like LSP, and Higgsino-like LSP cases.  To make our point
clear, we assume that the lightest stop can decay only into one of the
Bino-like, Wino-like, or Higgsino-like particles (including chargino).
Such a situation is realized when there is some hierarchy among the
parameters $M_1$, $M_2$, and $\mu$.

\begin{figure}
\hspace*{0.2cm}
\includegraphics[height=6.5cm]{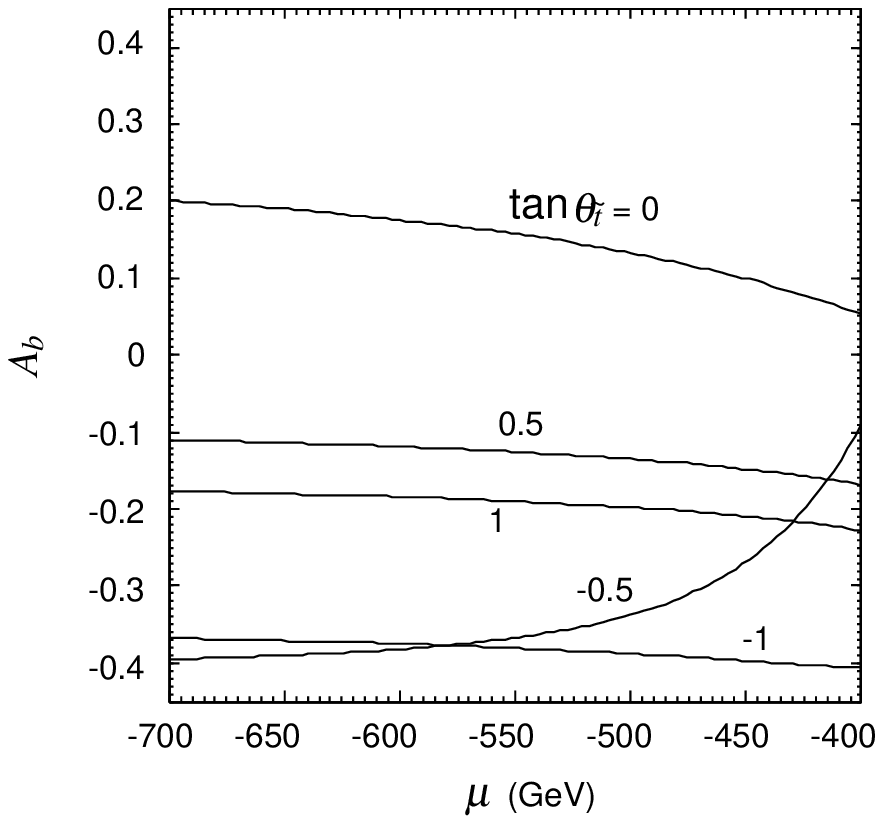}  
\hspace*{0.5cm}
\includegraphics[height=6.5cm]{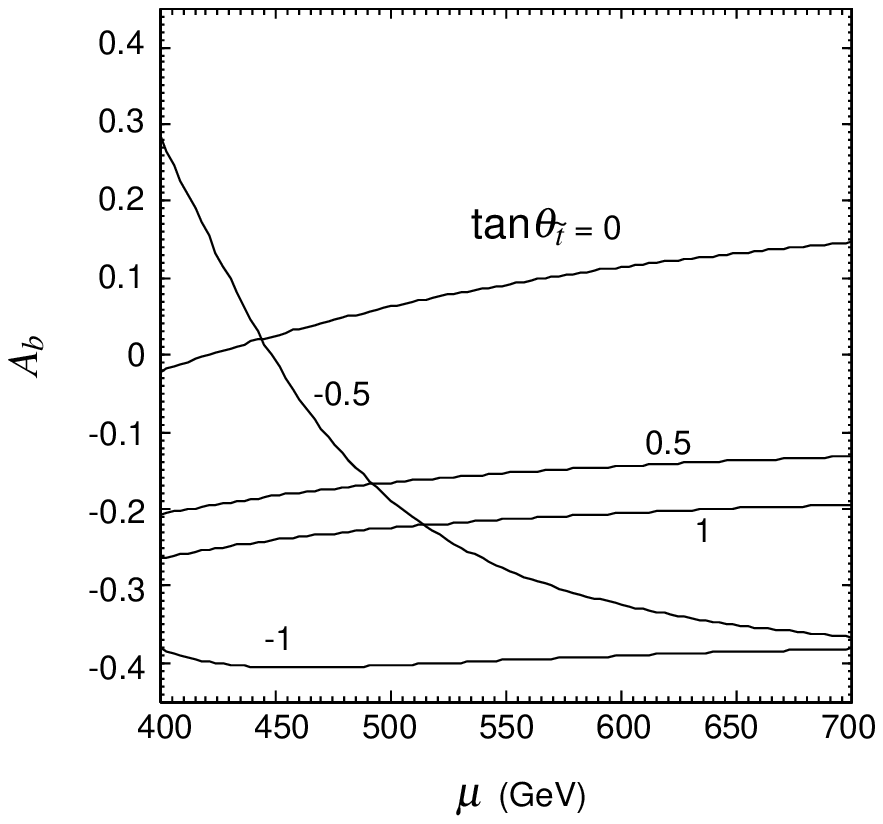}  
\caption{$A_b$ as a function of $\mu$-parameter
for $\tan \theta_{\tilde{t}}=-1$, $-0.5$, $0$, $0.5$, $1$.  We take
$\tan\beta=10$, $m_{\tilde{t}_1} = 400$ GeV, $m_{\chi_1^0} = 200$ GeV,
and $M_2=500$ GeV.}
\label{fig:Ab-mu(B)}
\end{figure}

\subsection{Bino-like LSP}
Even if the LSP is ``Bino-like,'' there can be sizable contaminations
of the Higgsino components in the LSP which affect the $f_L$- and
$f_R$-parameters.  It is instructive to consider the case $\langle
H_1\rangle,\langle H_2\rangle\ll |M_1|, |M_2|, |\mu|$ in order to
study effects of the Higgsino components.  
In this case, the mixing matrix elements
are approximately given by $[U_{\chi^0}]_{1\tilde{B}}\simeq 1$,
$[U_{\chi^0}]_{1\tilde{W}}\simeq 0$,
\begin{eqnarray}
[U_{\chi^0}]_{1\tilde{H}_1} \simeq
- \frac{m_Z \sin\theta_W (M_1\cos\beta + \mu\sin\beta)}
{\mu^2 - M_1^2},
\end{eqnarray}
and
\begin{eqnarray}
[U_{\chi^0}]_{1\tilde{H}_2} \simeq
\frac{m_Z \sin\theta_W (M_1\sin\beta + \mu\cos\beta)}
{\mu^2 - M_1^2},
\label{U-14}
\end{eqnarray}
where $\sin \theta_W$ is the weak mixing angle and $\tan \beta$ is the
ratio of the two vacuum expectation values of the Higgs fields, i.e.,
$\tan \beta = \langle H_2 \rangle / \langle H_1 \rangle$.  Thus, even if
$[U_{\chi^0}]_{1\tilde{B}}$ is close to 1,
$[U_{\chi^0}]_{1\tilde{H}_2}\sim O(0.1)$ is possible in a large
parameter space.  In addition, it is important to note that $y_t$ is
large and hence the Higgsino contributions to the coupling constants
$f_L$ and $f_R$ are enhanced.  Since $[U_{\chi^0}]_{1\tilde{H}_2}$ has
strong dependence on $\mu$, the $\mu$-parameter may be constrained from
the measurement of $A_b$.  $[U_{\chi^0}]_{1\tilde{H}_2}$ also depends on
$\tan\beta$ when $\tan\beta$ is small.  On the contrary, when
$\tan\beta$ is large, the $\tan\beta$ dependence becomes weak.

In Fig.\ \ref{fig:Ab-mu(B)}, we plot $A_b$ as a function of $\mu$ for
several values of $\tan\theta_{\tilde{t}}$.  As expected, $A_b$ varies
depending on the value of $\theta_{\tilde{t}}$.  More interestingly,
when $-0.5\lesssim\tan\theta_{\tilde{t}}\lesssim 0$, $\mu$ dependence
of $A_b$ becomes sizable.  This behavior can be understood as
follows.  As seen in Fig.\ \ref{fig:theta_vs_A}, $A_b$ becomes
sensitive to $f_L/f_R$ when $f_L/f_R\sim 1$.  In the limit
$|[U_{\chi^0}]_{1\tilde{H}_2}|\ll 1$, this relation is realized when
$\tan\theta_{\tilde{t}}=-\frac{1}{4}$ for the Bino-like LSP case.
Thus, when $\tan\theta_{\tilde{t}}\sim -\frac{1}{4}$, $A_b$ becomes
the most sensitive to $[U_{\chi^0}]_{1\tilde{H}_2}$ and hence $A_b$
acquires a strong dependence on $\mu$.  Take the line of
$\tan\theta_{\tilde{t}}=-0.5$ as an example, for negative $\mu$ (left
plot), $A_b$ reaches its asymptotic value when $\mu \sim -550$ GeV,
while for positive value of $\mu$ (right plot), $A_b$ remains
sensitive to $\mu$ up to a larger value of $\mu$.  This behavior can
be understood using Eq.~(\ref{U-14}).  With $M_1>0$ with $|\mu|$
fixed, $[U_{\chi^0}]_{1\tilde{H}_2}$ is more enhanced when $\mu>0$.
Thus, positive $\mu$ renders larger contamination of Higgsino
component in $\chi_1^0$, which in turn makes $\mu$ dependence of $A_b$
stronger.  Notice that the results with $M_1<0$ is obtained by the
previous results with opposite sign of $\mu$.

\begin{figure}
\hspace*{0.5cm}
\includegraphics[height=7cm]{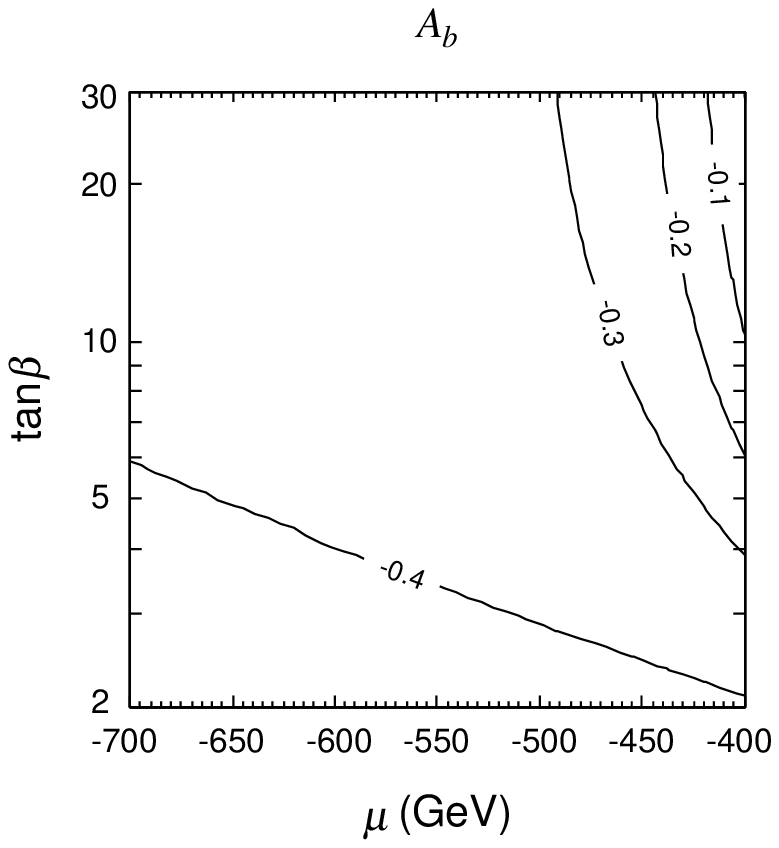}
\hspace*{0.7cm}
\includegraphics[height=7cm]{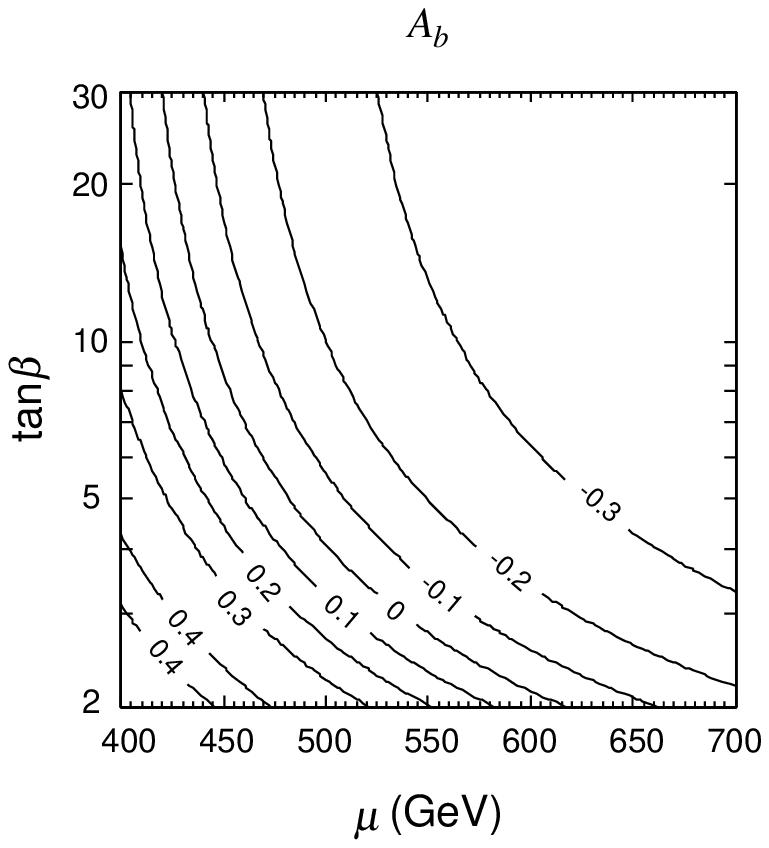}
\caption{Contour plot of $A_b$ on the $\mu$ vs.\ $\tan \beta$ plane
for Bino LSP case with $\tan \theta_{\tilde{t}} = -0.5$.  Stop and LSP
masses are taken to be $m_{\tilde{t}_1} = 400$ GeV and $m_{\chi_1^0} =
200$ GeV, respectively.  $M_2$ is taken to be $500\ {\rm GeV}$.}
\label{fig:Ab_in_mu-tanbeta(B)}
\end{figure}

In Fig.\ \ref{fig:Ab_in_mu-tanbeta(B)}, we plot contours of constant
$A_b$ on the $\mu$ vs.\ $\tan\beta$ plane for $m_{\tilde{t}_1}=400\ 
{\rm GeV}$, $m_{\chi_1^0}=200\ {\rm GeV}$, and
$\tan\theta_{\tilde{t}}=-0.5$.  With this choice of parameters,
$N_{\rm tot}\simeq 480$ for $L=100$ fb$^{-1}$ and hence we expect the
experimental uncertainty of $\delta A_b\simeq 0.1$.  Thus, quite a
severe constraint may be derived on the $\mu$ vs.\ $\tan\beta$
plane. For negative $\mu$, regions up to $-500$ GeV is sensitive to
$A_b$ measurements at a linear collider.  While for positive $\mu$,
the $\mu$ dependence becomes stronger in much wider region.
Especially for a fixed value of $\tan\beta$, the $\mu$-parameter may
be determined with an accuracy of $\delta\mu\sim 20\ {\rm GeV}$.

\subsection{Wino-like LSP}

If the LSP is the Wino-like neutralino,\footnote
{For the Wino-like LSP, $|M_2|<|M_1|$ is required.  This relation can
be realized, for example, in the minimal anomaly-mediated model
\cite{JHEP9812-027,NPB557-79,JHEP0004-009}.}
a similar analysis is possible.  Of course, in the Wino-like LSP case,
the lightest chargino is almost degenerate with the LSP and hence the
stop may also decay into the chargino and the bottom quark.  The mass
splitting between $\chi^\pm_1$ and $\chi^0_1$ is expected to be $O(100\
{\rm MeV})$ to a few GeV \cite{JHEP9812-027,PRL83-1731} and hence
$\chi^\pm_1$ decays into $\chi^0_1$ and soft lepton(s) or (light)
meson(s).  Thus, it is possible to distinguish the stop event with
$\tilde{t}_1\rightarrow t\chi^0_1$ and that with $\tilde{t}_1\rightarrow
b\chi^\pm_1$.  In addition, we calculated the branching ratio going into
top and the LSP, and found that $Br(\tilde{t}_1\rightarrow t\chi^0_1)$
becomes $O(0.1)$ in a large parameter space.  Thus, even if we cannot use all
of the stop events, it is still possible to do the same analysis as
before.

\begin{figure}
\hspace*{0.2cm}
\includegraphics[height=6.5cm]{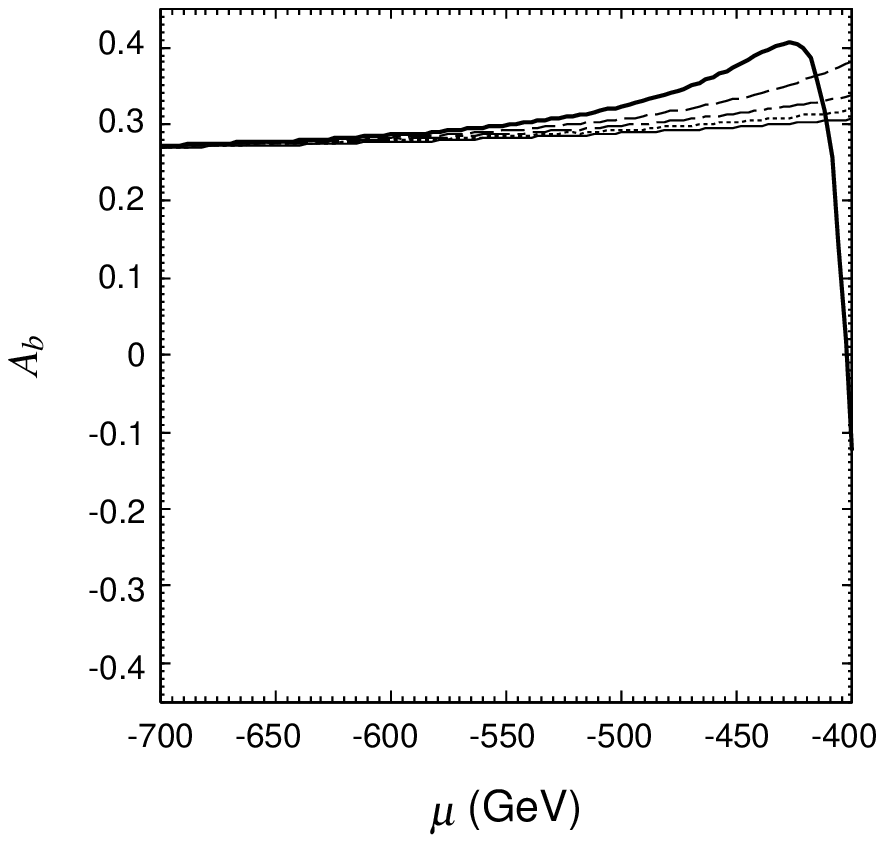}
\hspace*{0.5cm}
\includegraphics[height=6.5cm]{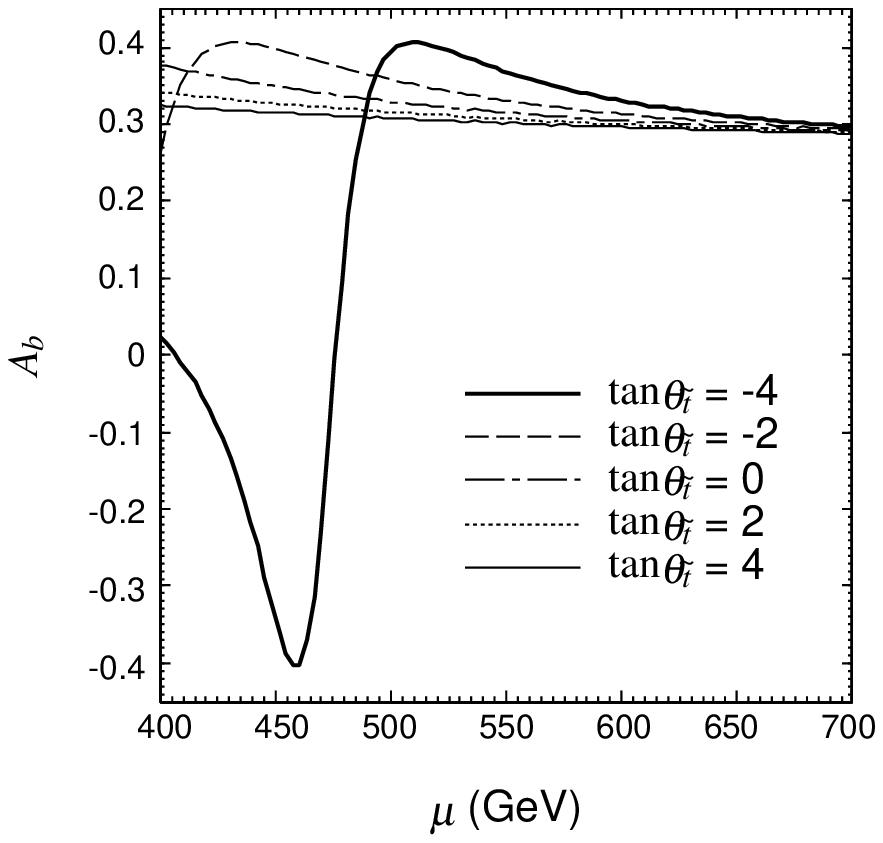}
\caption{$A_b$ as a function of $\mu$-parameter
for $\tan \theta_{\tilde{t}} = -4$ (thick line),
$-2$ (dashed line),
$0$ (dash-dotted line),
$2$ (dotted line), and
$4$ (solid line).
We take $\tan \beta = 10$, $m_{\tilde{t}_1} = 400$ GeV,
$m_{\chi_1^0} = 200$ GeV, and $M_1=$ 500 GeV.}
\label{fig:Ab-mu(W)}
\end{figure}

In Wino-like LSP scenario, the neutralino mixing matrix elements are 
approximately  given by 
$[U_{\chi^0}]_{1\tilde{W}}\simeq 1$, $[U_{\chi^0}]_{1\tilde{B}}\simeq
0$,
\begin{eqnarray}
[U_{\chi^0}]_{1\tilde{H}_1} \simeq
\frac{m_Z \cos\theta_W (M_2\cos\beta + \mu\sin\beta)}
{\mu^2 - M_2^2},
\end{eqnarray}
and 
\begin{eqnarray}
[U_{\chi^0}]_{1\tilde{H}_2} \simeq -
\frac{m_Z \cos\theta_W (M_2\sin\beta + \mu\cos\beta )}
{\mu^2 - M_2^2}.
\label{eq:U-14-wino}
\end{eqnarray}
Similarly to the Bino-LSP case, contamination of the Higgsino
component in the LSP also induces $\mu$ dependence of $A_b$.  The
ratio $f_L/f_R$ is now given by
\begin{equation}
\frac{f_L}{f_R}\simeq -\frac{1}{\sqrt{2}}
\frac{g_2}{y_t [U_{\chi^0}]_{1H_2}}+\tan\theta_{\tilde{t}}. 
\end{equation}
Since $[U_{\chi^0}]_{1\tilde{H}_2}$ is usually small and negative (we
have adopted the convention that $M_2>0$), the relation $f_L/f_R\sim 1$
requires a negative $\tan\theta_{\tilde{t}}$ with
$|\tan\theta_{\tilde{t}}| \gg 1$.  This is shown in Fig.\
\ref{fig:Ab-mu(W)}, where we plot $A_b$ as a function of $\mu$-parameter
for the Wino-like LSP case.  The parameters are chosen to be $\tan \beta
= 10$, $m_{\tilde{t}_1} = 400$ GeV, $m_{\chi_1^0} = 200$ GeV, and $M_1=$
500 GeV.  $A_b$ becomes sensitive to the $\mu$-parameter for
$\tan\theta_{\tilde{t}}\sim -4$.  For smaller values of negative
$\tan\theta_{\tilde{t}}$, the curve shifts to the right (left) for
$\mu>0$ ($\mu<0$).  Such a behavior can be explained by using Eq.\
(\ref{eq:U-14-wino}); a larger value of $|\mu|$ is needed to compensate
smaller value of (negative) $\tan\theta_{\tilde{t}}$.  The different
behaviors for the positive and negative $\mu$ cases are due to the same
reason as explained in the Bino-like LSP case.  The dependence on $M_1$
is rather weak for large values of $M_1$.

\begin{figure}
\hspace*{0.5cm}
\includegraphics[height=7cm]{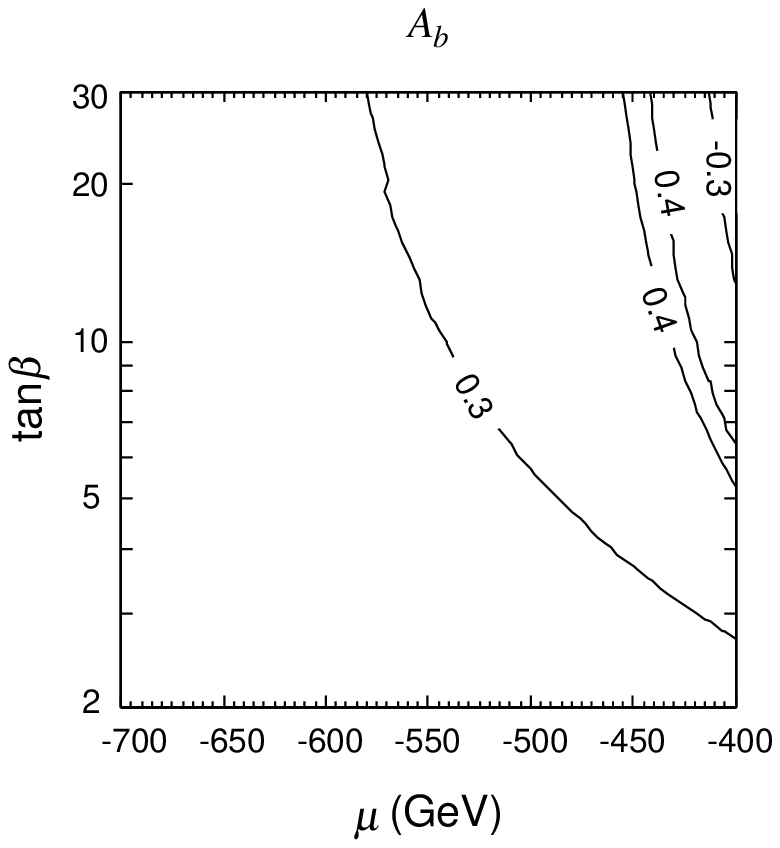}
\hspace*{0.7cm}
\includegraphics[height=7cm]{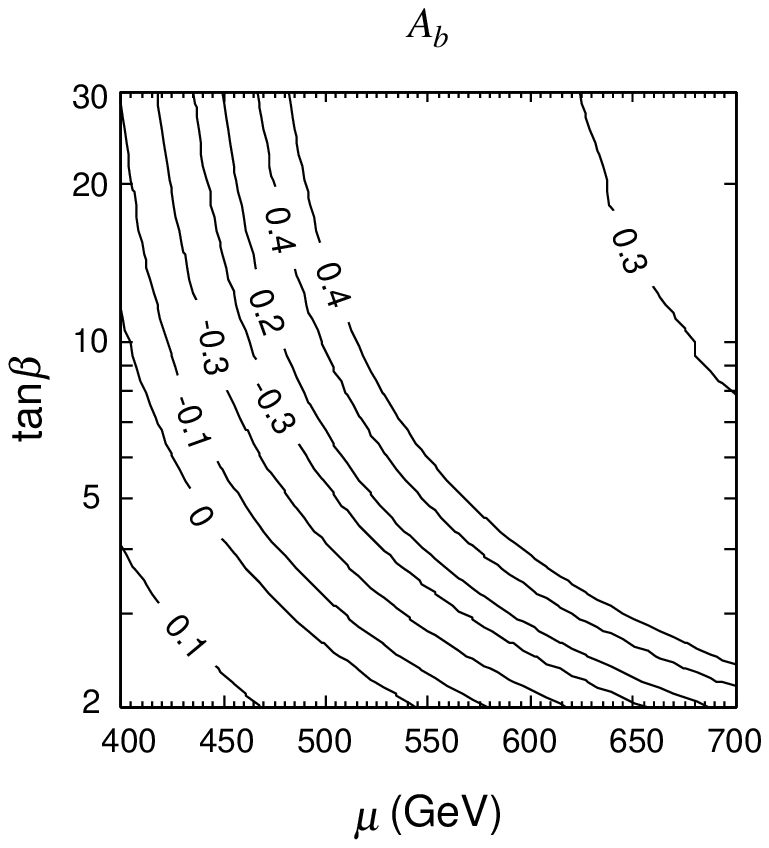}
\caption{Contour plot of $A_b$ on the $\mu$ vs.\ $\tan \beta$ plane
for Wino LSP case with $\tan \theta_{\tilde{t}} = -4$.  Stop and LSP
masses are taken to be $m_{\tilde{t}_1} = 400$ GeV and $m_{\chi_1^0} =
200$ GeV, respectively.  $M_1$ is taken to be 500 GeV.  }
\label{fig:Ab_in_mu-tanbeta(W)}
\end{figure}

In Fig.\ \ref{fig:Ab_in_mu-tanbeta(W)}, we plot the contours of
constant $A_b$ on the $\mu$ vs.\ $\tan\beta$ plane for the Wino-like
LSP case with $\tan\theta_{\tilde{t}}=-4$.  The constraint on $\mu$
and $\tan\beta$ from the $A_b$ measurement is obtained in the same way
as the Bino LSP case but the statistical error of $A_b$ depends on the
branching ratio $Br(\tilde{t}_1\to t\chi_1^0)$ since $\delta
A_b\propto 1/\sqrt{Br(\tilde{t}_1 \to t \chi_1^0)}$.  Although the
number of events is reduced, we still have a possibility of
constraining the values of $\mu$ and $\tan\beta$ as one can see in
Fig.\ \ref{fig:Ab_in_mu-tanbeta(W)}.  Interestingly, very steep
dependence on $\mu$ is realized in some parameter region, as shown in
Fig.\ \ref{fig:Ab_in_mu-tanbeta(W)}.  

\begin{figure}
\hspace*{0.5cm}
\includegraphics[height=7cm]{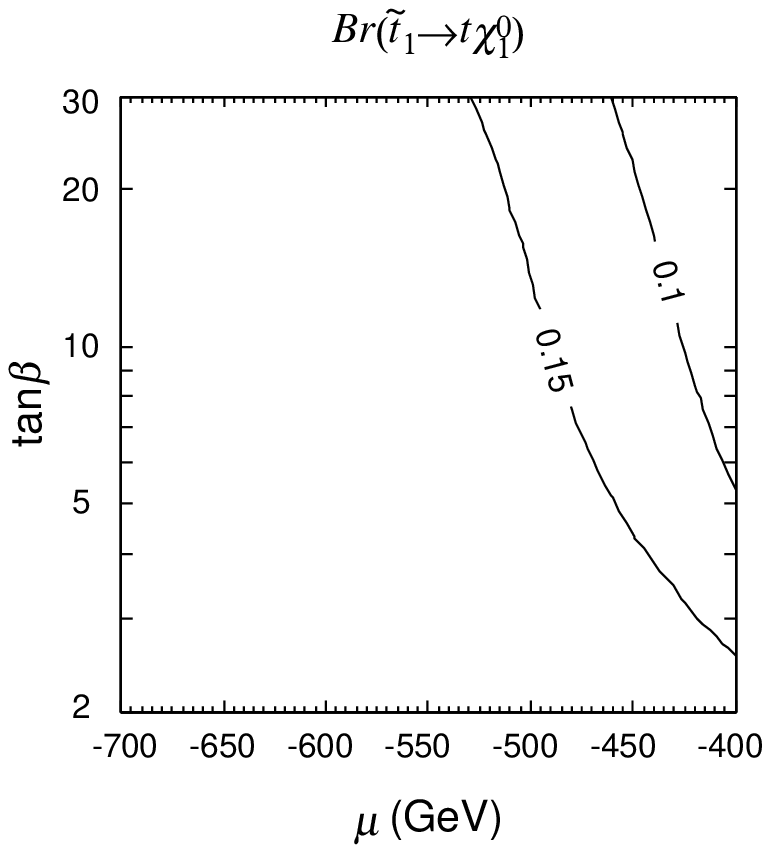}
\hspace*{0.7cm}
\includegraphics[height=7cm]{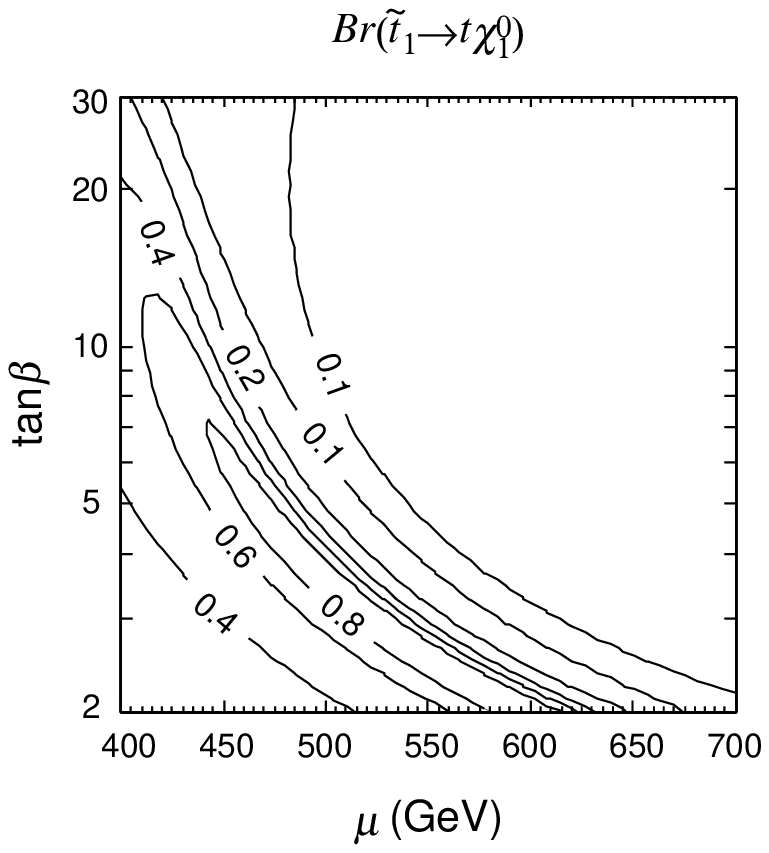}
\caption{$Br(\tilde{t}_1 \to t \chi_1^0)$ on the $\mu$ vs.\ $\tan \beta$
plane for Wino LSP case with $\tan\theta_{\tilde{t}} = -4$.  Stop and
LSP masses are taken to be $m_{\tilde{t}_1} = 400$ GeV and
$m_{\chi_1^0} = 200$ GeV, respectively, and $M_1$ is taken to be 500
GeV.  }
\label{fig:br_in_mu-tanbeta(W)}
\end{figure}

One should also note that we can use the branching ratio
$Br(\tilde{t}_1 \to t \chi_1^0)$ as another observable in the Wino LSP
case.  In Fig.\ \ref{fig:br_in_mu-tanbeta(W)}, we plot the contours of
$Br(\tilde{t}_1 \to t \chi_1^0)$ on the $\mu$ vs.\ $\tan\beta$ plane
for $\tan\theta_{\tilde{t}}=-4$.  As one can see, $Br(\tilde{t}_1\to t
\chi_1^0)$ has sizable dependence on $\mu$ and $\tan\beta$, so that
independent information on these parameters can be obtained.

\subsection{Higgsino-like LSP}

Finally we comment on the Higgsino-like LSP case.  In this case, $f_L$
and $f_R$ are both dominated by the top-Yukawa contribution and hence
the ratio $f_L/f_R$ becomes $\sim\tan\theta_{\tilde{t}}$.  In this case,
$A_b$ is insensitive to the gaugino masses and it is difficult to derive
a constraint on $M_1$ and $M_2$.  In the Higgsino-like LSP case,
however, mass splitting between $\chi^0_1$ and $\chi^\pm_1$ is typically
$\sim 10$ GeV or larger \cite{Drees:1996pk} and hence direct studies of
the chargino will be possible.  In addition, the second lightest
neutralino is also quite degenerate with $\chi^0_1$ and hence is likely
to be kinematically accessible.  Thus, in this case, rather than
studying the production and the decay of $\tilde{t}_1$, one should
better study the productions and decays of $\chi^\pm_1$ and $\chi^0_2$
to understand the properties of the neutralinos and charginos
\cite{Feng:1995zd}.

\section{Conclusions}
\label{sec:conclusions}

We have studied the lightest stop pair production and the stop decay
mode of $\tilde{t}_1 \to t \chi_1^0$, followed by $t \to b W$ at future
$e^+e^-$ linear colliders.  From the naturalness point of view, stop
should not be very heavy since it has a strong Yukawa interaction with
the Higgs particle.  Large top Yukawa coupling constant also makes one
of the stop mass eigenstates light through left-right mixing in the stop
mass matrix.  Therefore it is reasonable to expect that linear colliders
can produce the lightest stops in pair directly.  If so, due to the
large top Yukawa coupling constant, rich information on the model
parameters is obtained by the analysis of the stop production and decay
processes.

We discuss that information on $m_{\tilde{t}_1}$, $\theta_{\tilde{t}}$
and $m_{\chi_1^0}$ can be obtained through the measurement of the
lightest stop pair production cross section using polarized electron
beam and by studying the end-point energy of the top quark. We also show
that the analysis of the angular distribution of the $b$-jets is useful
to extract information on the neutralino sector such as neutralino
mixing.  We found a parameter region where it is possible to give a
strong constraint on a combination of the $\mu$-parameter and
$\tan\beta$ assuming $\theta_{\tilde{t}}$, $m_{\tilde{t}_1}$, and
$m_{\chi^0_1}$ are known.  Even if we cannot obtain a strong constraint
on $\mu$ or $\tan\beta$, the analysis of $A_b$ is important since the
value of $f_L/f_R$ varies depending on what the LSP is.  For example,
let us consider the case where there is no chargino with a nearly
degenerated mass with the LSP.  In such a case, one might naively expect
that the LSP is Bino-like.  The stop analysis we proposed provides
independent information on the properties of the LSP.  Such a test is
significant since the value of $A_b$ may deviate from that in the
Bino-LSP case if there is an extra neutralino such as the superpartner
of the singlet Higgs boson.

\section*{Acknowledgments}
The authors would like to thank K. Fujii and Y. Sumino for useful
discussions and comments.  T.M. and S.S. thank the Aspen Center for
Physics where a part of this work has been done.  The work of R.K. is
supported by JSPS.  The work of T.M. is supported by the Grant-in-Aid
for Scientific Research from the Ministry of Education, Science,
Sports, and Culture of Japan, No.\ 12047201 and No.\ 13740138.  The
work of S.S is supported by the DOE under grant DE-FG03-92-ER-40701
and by the John A. McCone Fellowship.

\end{document}